\documentclass[iop]{emulateapj}
\usepackage{ulem}
\usepackage{color}

\newcommand{\msun}{$M_\sun$}

\newcommand{\hh}{$H_{160}$}
\newcommand{\jh}{$J_{125}-H_{160}$}
\newcommand{\hone}{$H_{160}-[3.6]$}

\shorttitle{Spectral energy distributions at $z\sim8$}
\shortauthors{I. Labb\'e et al.}

\begin{document}

\def\figstamps{
\begin{figure}
\begin{center}
\includegraphics[width=8.9cm]{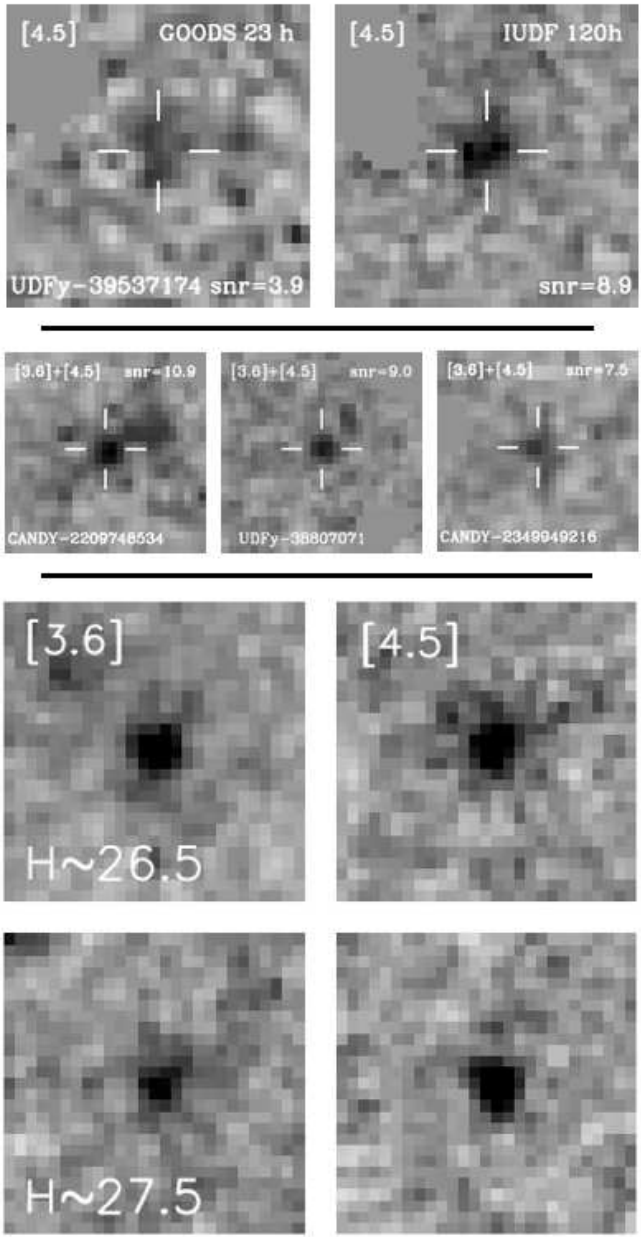}
\end{center}
\leavevmode			
\vspace{-0.7cm}
\caption{
({\it top panels}) A comparison of Spitzer/IRAC $[4.5]$ band images between GOODS 
(23h exposure time) and the new IUDF observations (120h) for $Y_{105}-$dropout 
galaxy UDFy-3953714 \citep{Bouwens11c} at $z\sim8$. Nearby foreground sources 
have been subtracted. With the new IUDF IRAC data the object
is clearly detected, even in the shallower IRAC [4.5] micron band. 
({\it middle panels}) Summed $[3.6] + [4.5]$ micron images 
of several IRAC detected $Y_{105}-$dropouts from our sample. 
({\it bottom panels}) Median stacked IRAC images of $z\sim8$ $Y$-dropout sources, 
grouped in $\sim1-$mag bins centered on $H_{160}\approx26.5,27.5$, containing
10 and 23 galaxies, respectively. Importantly, the stacks show significant 
detections at $[3.6]$, { which at $z\sim8$ is only moderately affected
by emission lines}. Image panels are shown in inverted grayscale and are 
$10\arcsec\times10\arcsec$.
\label{figstamps}}
\end{figure}
}

\def\figstack{
\begin{figure*}
\epsscale{1.0}
\centering
$$ $$
\includegraphics[width=17cm,bb=54 330 876 1100]{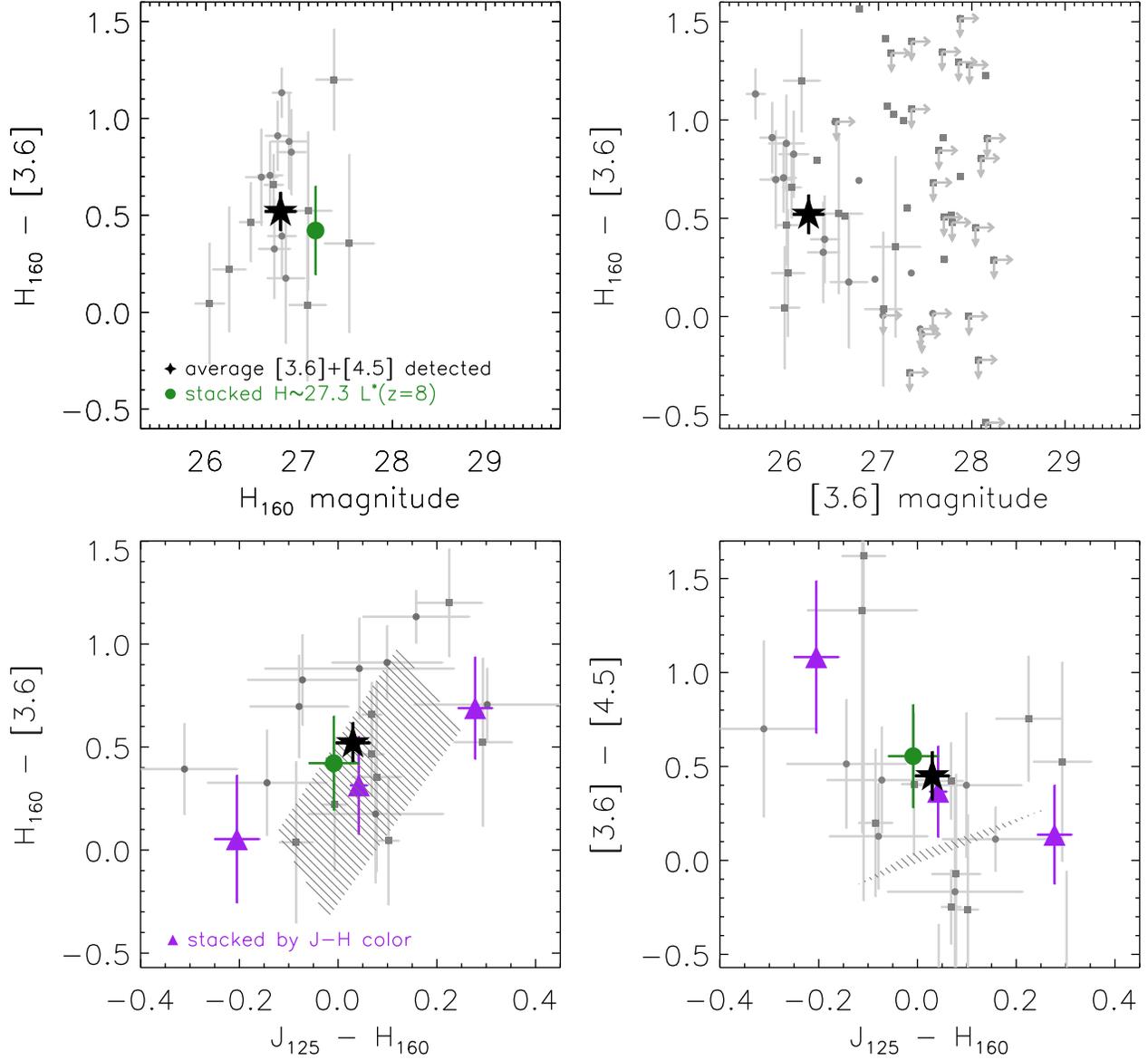}
$$ $$
\leavevmode
\vspace{-1.5cm}
\caption{A first exploration of the observed IRAC properties of $z\sim8$
galaxies in our fields. Galaxies directly detected at $>3\sigma$ in {IRAC bands $[3.6]+[4.5]$ are shown by gray 
points (squares for ultradeep IRAC from IUDF, circles for wide-field IRAC from GOODS)}. 
The averages of IRAC detected galaxies are indicated by filled stars. {Stacked measurements in a 1-magnitude 
bin centered at $H=27.3$ ($\sim L^*_{UV}(z=8)$, including IRAC undetected sources, are shown in green.
Stacking measurements of $H \leq 28$ galaxies in bins of \jh\ colors are shown in purple.}
{\it Top left panel:} the \hone\ versus \hh\ diagram.
{\it Top right panel:} the observed $H-[3.6]$ versus $[3.6]$
magnitude. Sources with {SNR$([3.6]+[4.5]) < 3\sigma$} are also shown (upper
limits are $1\sigma$ in $[3.6]$). The deepest data over the HUDF reach 
$[3.6]\sim28$ mag AB ($1\sigma$). {\it Lower left panel:} the observed $H-[3.6]$ 
versus \jh\ is in the range predicted by dust-reddened
\citet[][BC03]{bc03} stellar population synthesis models, with
some evidence for a correlation of $H-[3.6]$ with \jh\ . The hatched region shows models with a  
formation redshift $z=10$ in the redshift range $7.3 < z < 8.3$ and 
with Calzetti et al. (2000) reddening $E(B-V) < 0.2$.
{\it Lower right panel:}  in contrast, the  observed $[3.6]-[4.5]$ colors 
are on average $\sim0.5$ mag redder than the BC03 models 
(rest-frame $B_{4000} - V_{5000}$), in particular at bluer \jh\ colors.
Likely, the $[4.5]$ flux is boosted by the contribution of the emission lines 
[\ion{O}{3}]$\lambda\lambda4959,5007$ and $H\beta$.
\label{figstack}}
\end{figure*}  
}

\def\figsed{
\begin{figure}		
\epsscale{1}
\centering
$$
\includegraphics[width=9.1cm,bb=74 360 357 586]{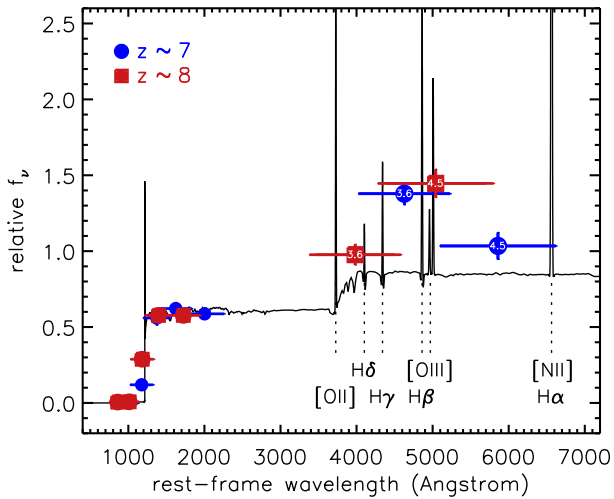} 
$$
\leavevmode
\vspace{-0.3cm}
\caption{
 A comparison of the average rest-frame SEDs  
 { of IRAC detected galaxies at $z\sim7$ and $z\sim8$  from the average HST/ACS, HST/WFC3, 
 and Spitzer/IRAC fluxes. The IRAC $[3.6]$ and $[4.5]$ fluxes are indicated. The SEDs are in
 units  of $f\nu$ (arbitrary scaling).
A young star forming stellar population model with emission lines is shown in black.
 The observed $[3.6]-[4.5]$ colors at $z\sim7$ and $z\sim8$ are substantially different, despite the short 
 time elapsed between these epochs (about 130Myr). Combined, the rest-frame SEDs suggest a clear
 flux excess at $\sim5000$\AA, shifting from $[3.6]$ at $z\sim7$ to $[4.5]$ at $z\sim8$, likely 
 due to a contribution from strong
[\ion{O}{3}]${4959,5007}$ and $H\beta$ emission lines. }
\label{figsed}}
\end{figure}
}

\def\figcolz{
\begin{figure}		
\epsscale{1}
\centering
\includegraphics[width=9.0cm]{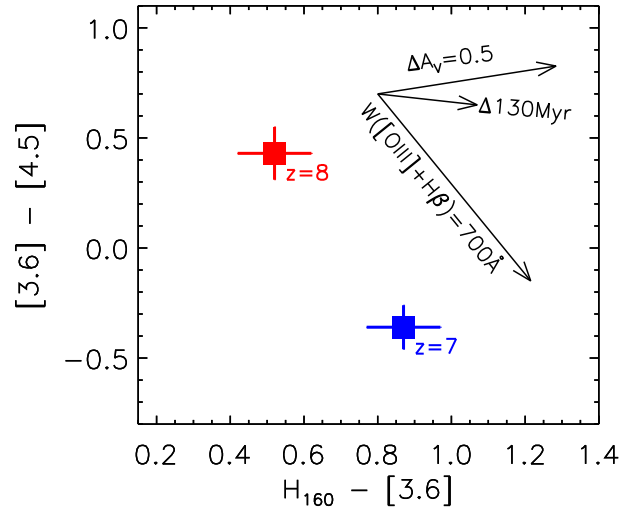} 
\leavevmode
\vspace{-0.3cm}
\caption{
The $H-[3.6]$ versus $[3.6]-[4.5]$ colors of our samples of 
$z\sim7$ and $z\sim8$ galaxies. The $[3.6]-[4.5]$ becomes $\approx0.8$ mag 
bluer from $z\sim8$ to $z\sim7$ while the $H-[3.6]$ colors become $\approx0.4$ mag redder.
The arrows show the effect of 1) increasing dust obscuration by $\Delta A_V=0.5$ 
between the two epochs, 2) changing the stellar population age by 130Myr 
(assuming CSF since $z=10$), and 3) strong [\ion{O}{3}]$_{4959,5007}$+$H\beta$ 
emission moving from $[4.5]$ at $z\sim8$ into $[3.6]$ at $z\sim7$.
Emission lines provide the most probable explanation for the observed change.}
\end{figure}
}

\def\figewz{
\begin{figure}		
\epsscale{1}
\centering
\includegraphics[width=8cm,bb=54 320 408 666]{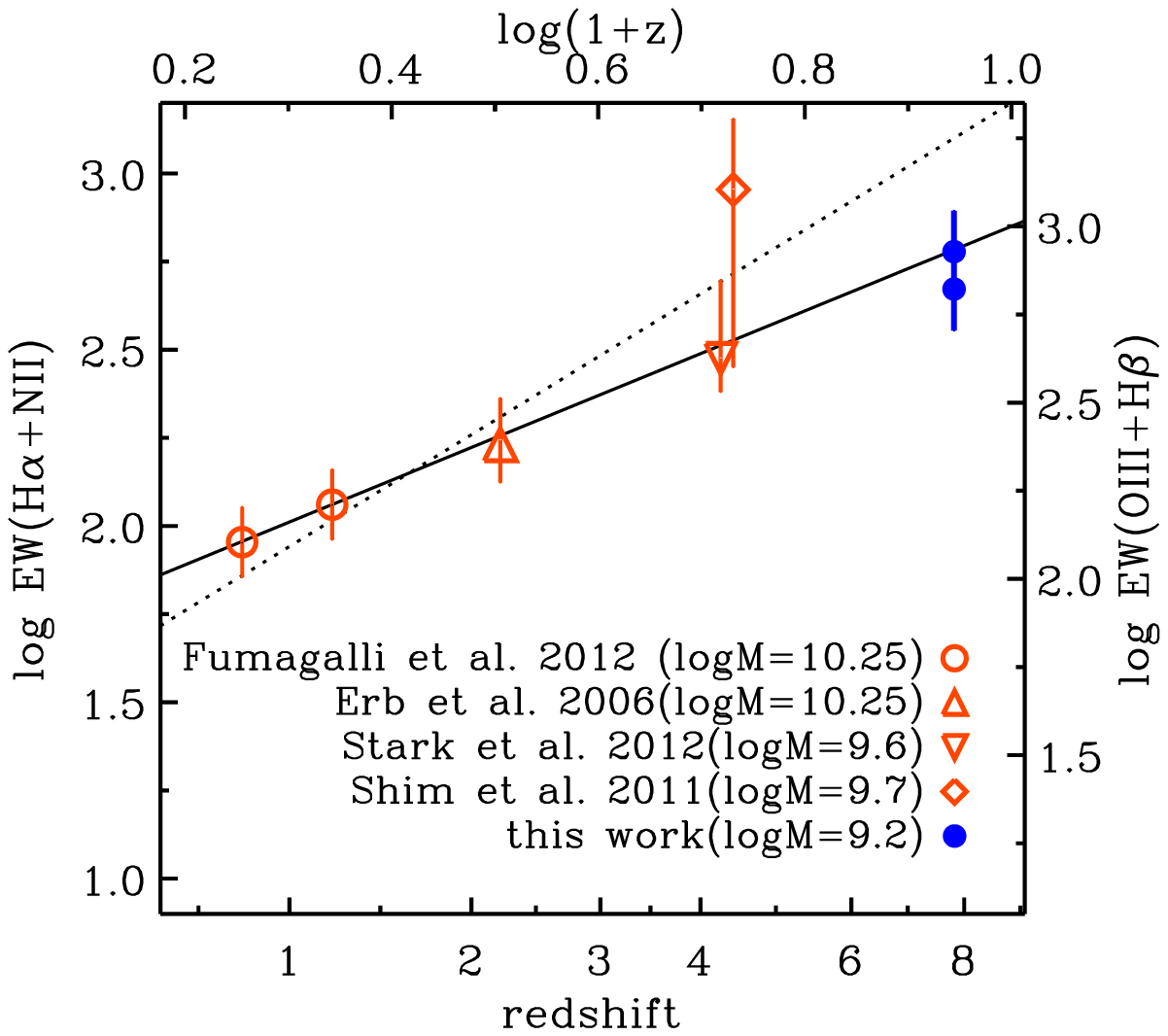} 
\leavevmode
\vspace{-0.65cm}
\caption{
Comparison of the rest-frame emission line EW of star forming galaxies 
with redshift. Red points indicate
estimates of the mean $H\alpha$+[NII] EW with redshift, using either
near-infrared spectra (Fumagalli et al. 2012, Erb et al. 2006) or 
emission line strengths inferred from excess emission in the IRAC $[3.6]$ filter
of spectroscopically confirmed galaxies at $z\sim4-5$. The broadband
derived $[OIII]+H\beta$ EW at $z\sim8$ from this paper are shown by the blue 
solid symbols. They are placed on the same scale assuming the relative 
emission line strengths of Anders \& Fritze-v.Alvensleben (2003)
for sub-Solar $0.2Z_{\sun}$ (lower value) and Solar (higher value). 
The dotted line shows the EW evolution $\propto(1 + z)^{1.8}$ derived by 
Fumagalli et al. (2012) for star forming galaxies with stellar 
mass $10.0 <\log(M/M\odot) < 10.5$ at $0<z<2$ extrapolated to $z=8$.
The $z\sim8$ point provides additional evidence for an increasing
contribution of emission lines towards higher redshifts. {The solid line
shows the best-fit linear relation to all data at $1<z<8$: W$_{H\alpha}\propto(1+z)^{1.2\pm0.25}$.}}
\end{figure}
}

\title{ THE SPECTRAL ENERGY DISTRIBUTIONS OF $z\sim8$ GALAXIES FROM THE IRAC
ULTRA DEEP FIELDS: EMISSION LINES, STELLAR MASSES, AND SPECIFIC STAR
FORMATION RATES AT 650 MYR
\altaffilmark{1}}

\author{I. Labb\'e\altaffilmark{2},
P. A. Oesch\altaffilmark{3,4},
R. J. Bouwens\altaffilmark{2}, 
G. D. Illingworth\altaffilmark{3}, 
D. Magee\altaffilmark{3}
V. Gonz\'alez\altaffilmark{3}, 
C. M. Carollo\altaffilmark{5}, 
M. Franx\altaffilmark{2}, 
M. Trenti\altaffilmark{6}, 
P. G. van Dokkum\altaffilmark{7}
M. Stiavelli\altaffilmark{8}
}

\altaffiltext{1}{Based on observations made with the NASA/ESA Hubble Space
Telescope, which is operated by the Association of Universities for
Research in Astronomy, Inc., under NASA contract NAS 5-26555. These
observations are associated with programs \#11563, 9797. Based on observations 
with the {\em Spitzer Space Telescope}, which is operated by the Jet 
Propulsion Laboratory, California Institute of Technology under 
NASA contract 1407.  Support for this work was provided by NASA through contract 
125790 issued by JPL/Caltech. Based on service mode observations collected at 
the European Southern Observatory, Paranal, Chile (ESO Program 073.A-0764A).
Based on data gathered with the 6.5 meter Magellan Telescopes located at Las Campanas 
Observatory, Chile.}
\altaffiltext{2}{Leiden Observatory, Leiden University, NL-2300 RA Leiden, Netherlands}
\altaffiltext{5}{UCO/Lick Observatory, University of California, Santa Cruz, CA 95064}
\altaffiltext{4}{Hubble Fellow}
\altaffiltext{5}{Institute for Astronomy, ETH Zurich, 8092 Zurich, Switzerland}
\altaffiltext{6}{Kavli Institute for Cosmology and Institute of Astronomy, University of Cambridge, United Kingdom}
\altaffiltext{7}{Department of Astronomy, Yale University, New Haven, CT 06520}
\altaffiltext{8}{Space Telescope Science Institute, Baltimore, MD 21218, United States}
 
\begin{abstract}	 
Using new ultradeep Spitzer/IRAC photometry from the IRAC Ultradeep Field 
program (IUDF), we investigate the stellar populations of a sample of 63 $Y-$dropout 
galaxy candidates at $z\sim8$, only 650Myr after the Big Bang. 
The sources are selected from  HST/ACS+WFC3/IR data over the Hubble Ultra Deep 
Field (HUDF), two HUDF parallel fields, and wide area data over the 
CANDELS/GOODS-South. The new Spitzer/IRAC data increase the coverage in $[3.6]$ 
and $[4.5]$ to $\sim120$h over the HUDF reaching depths of $\sim28$ (AB,$1\sigma$). 
The improved depth and inclusion of brighter candidates result in direct $\geq3\sigma$ 
IRAC detections of 20/63 sources, of which 11/63 are detected at $\geq5\sigma$. 
{The average $[3.6]-[4.5]$ colors of IRAC detected galaxies at $z\sim8$ are markedly
redder than those at $z\sim7$, observed only 130Myr later.}
The simplest explanation is that we witness strong rest-frame optical emission lines 
(in particular [\ion{O}{3}]$\lambda\lambda4959,5007+H\beta$) moving through the IRAC bandpasses 
with redshift. Assuming that the average rest-frame spectrum is the same at both $z\sim7$ 
and $z\sim8$ we estimate a rest-frame equivalent width of 
{W$_{[OIII]\lambda\lambda4959,5007+H\beta}=670^{+260}_{-170}$\AA\ contributing 
$0.56^{+0.16}_{-0.11}$ mag to the $[4.5]$ filter at $z\sim8$. The corresponding
W$_{H\alpha}=430^{+160}_{-110}$\AA\ implies an average specific star formation rate of 
$sSFR=11_{-5}^{+11}$~Gyr$^{-1}$ and a stellar population age of $100_{-50}^{+100}$ Myr.
Correcting the spectral energy distribution for the contribution of emission 
lines lowers the average best-fit stellar masses and mass-to-light ratios by $\sim3\times$, 
decreasing the integrated stellar 
mass density to $\rho^*(z=8,M_{UV}<-18)=0.6^{+0.4}_{-0.3}\times10^6~M_\sun~$Mpc$^{-3}$. } %update
\end{abstract}

\section{Introduction}

Deep imaging of high redshift $z>6$ galaxies at mid-infrared wavelengths 
with the InfraRed Array Camera (IRAC; \citealt{Fa04}) on {\it Spitzer} has
led to the surprising discovery that $z\geq6$ galaxies have 
red $H-[3.6]\sim0.5$ colors. This has been taken as evidence of
substantial stellar masses $\sim10^{9}-10^{10}M_\sun$ \citep{Ey05,Ya06,St09} 
and  stellar ages ($>300$Myr) \citep{La06,Go10,La10a,La10b}.
Early studies also suggested that the specific star formation rate 
at fixed stellar mass was nearly flat at $3<z<8$, in apparent 
disagreement with the strongly increasing specific inflow rates 
of baryons predicted by galaxy formation models  (e.g., Neistein \& Dekel 2008; 
Dav\'e et al. 2011, Weinmann et al. 2011).

A key uncertainty in the analysis of the photometry is how much 
rest-frame optical emission lines contribute to the broadband fluxes. 
If emission lines are very strong, then the stellar masses previously
derived by fitting stellar population models without lines would be 
biased high, and the specific star formation rate (sSFR) would be biased
low (e.g., \citealt{La10b}, \citealt{Sch10}).  To address this issue, 
 emission lines have been included in some stellar population 
models (e.g., Schaerer et al. 2010, 2012, de Barros et al. 2012, 
Yan et al. 2012).  As model predictions for emission lines are very 
uncertain empirical approaches are complementary and necessary. 

Fumagalli et al. (2012) studied the evolution of the 
$H\alpha$ equivalent width at $0<z<2$ using near-IR spectra and 
predict very large values at $z\sim8$. If the emission lines are indeed
as strong as argued, then deep photometry should be able
to reveal the effect of the lines moving through the IRAC passbands 
with  redshift. This technique has been applied successfully 
by looking for excess flux in the $[3.6]-$band due to $H\alpha$ 
at $z\sim4.5$ (e.g. Shim et al. 2011, Stark et al. 2013) and by 
comparing the average observed IRAC colors over $4<z<6$ (Gonzalez
et al. 2012). Nevertheless, the current situation is that the 
contribution of emission lines at $4<z<8$ is still poorly known.

\figstamps

Direct estimates at $z\sim8$ would provide the best constraints yet 
on the evolution of emission line strengths and specific star
formation rate to the highest redshifts. Fortunately, at redshift $z\sim8$ 
a relatively clean test is possible for the strength of the strongest
lines [\ion{O}{3}]${\lambda\lambda4959,5007}$+$H\beta$, because these
lie isolated in the $[4.5]$ filter. Then main challenge is the extreme 
faintness of the sources.

In this Letter, we use the largest sample of $z\sim8$ galaxy
candidates in combination with newly acquired ultradeep IRAC
data from the IRAC Ultra Deep Field program (IUDF; PI Labb\'e) 
to study their colors, SEDs, the contribution of emission lines,
and to derive emission line corrected stellar masses and specific
star formation rates. Throughout this paper, we assume an 
$\Omega_M=0.3,\Omega_\Lambda=0.7$ cosmology with $H_0=70$~km~s$^{-1}$Mpc$^{-1}$. 
Magnitudes are in the AB photometric system \citep{Ok83}.

\section{Data}

The data analyzed here consist of ultradeep WFC3/IR imaging
from the HUDF09 program (GO 11563: PI Illingworth) over 
the HUDF and two nearby fields HUDF09-1 and HUDF09-2, supplemented
with deep WFC3/IR data observations from the Early Release Science 
program (GO 11359: PI O'Connell) and the Multi-Cycle Treasury program 
CANDELS (PI: Faber/Ferguson; Grogin et al. 2011; Koekemoer et al. 2011) 
over the GOODS-South.

We use new ultradeep Spitzer/IRAC imaging from the IRAC Ultradeep 
Field program (IUDF; PI Labb\'e, PID 70145), a 262 hour Spitzer warm mission
program at [3.6] and [4.5] micron (Labb\'e et al. in preparation). 
This survey increases the exposure time over the HUDF, HUDF09-1 and 
HUDF09-2 fields from $12-46$ hours to $\sim$120 hours, $\sim$50 hours, 
and $80-120$ hours, 
respectively. For the wider GOODS area we use the 23$-$46 hour deep IRAC 
coverage of GOODS (M. Dickinson et al. in preparation).

\figstack

Our primary sample consists of 60 $Y-$dropout galaxies at $z\sim8$ 
selected by \citet{Bouwens11c} over the HUDF09 and ERS fields and
16 brighter $Y-$dropouts selected by \citet{Oesch12} over the 
CANDELS GOODS-South area. \footnote{{The ERS data uses the $Y_{098}$ 
filter versus the $Y_{105}$ filter in CANDELS+HUDF. However, the 
criteria from Bouwens et al. 2011 and Oesch et al 2012 were chosen in
such a way that the samples would have a very similar redshift
distribution.}}

We derive new IRAC photometry of all 76 sources following the procedure 
of \citet{La10a,La10b}. Briefly, we subtract nearby foreground sources
based on their HST image profiles and determine local backgrounds
(see e.g, Gonzalez et al 2011). 
Then we perform aperture photometry through $2\arcsec$ diameter 
apertures on the cleaned images and correct the flux for light outside
the aperture using a profile derived from nearby stars ($2.4\times$ in 
[3.6] and $2.5\times$ in [4.5]). Details of the photometry are presented 
in Labb\'e et al. (in preparation). We exclude 13 sources for which 
clean subtraction was not possible due to the proximity of very bright 
foreground sources, leaving a final sample of 63 $Y-$dropouts.

Figure~1 presents image stamps of IRAC$-$detected $Y-$dropout galaxies. 
A direct comparison to earlier GOODS observations demonstrates the clear 
improvement in sensitivity with the new IRAC Ultradeep Field data. 

We also include fainter, IRAC undetected, galaxies in the analysis by 
stacking $15\arcsec\times15\arcsec$ image stamps centered on the sources. 
On the stacks we subtract any residual background in a 
concentric annulus $5\arcsec<r<7.5\arcsec$. We compare between average 
and median combination to make sure that average results are not driven 
by outliers and derive uncertainties by bootstrap resampling.

\figsed

\section{Observed properties}

To explore the observed properties of $z\sim8$ galaxies, we first present color$-$magnitude
and color$-$color diagrams which contain diagnostic information about their stellar populations. 
Figure~2 ({\it left panel}) shows the 
\hone\ versus \hh\ color-magnitude diagram.  The improved IRAC depth and inclusion of brighter 
candidates result in a direct detection with signal-to-noise ratio SNR([3.6]+[4.5])$ \geq 3$
of 20/63 sources (32\%), of which 11 (17\%) are detected at $\geq5\sigma$.

{ The average colors of IRAC detected ($\geq 3\sigma$) galaxies are
$<H-[3.6]>=0.52\pm0.09$, with an intrinsic scatter of $\sigma(H-[3.6])=0.40\pm0.12$
(photometric uncertainties subtracted in quadrature), and $<[3.6]-[4.5]>=0.43\pm0.11$
with $\sigma(H-[3.6])=0.55\pm0.15$. Stacking fainter $H\sim27.3$ ($\sim L^*(z=8)$) galaxies, including IRAC undetected sources, 
shows similar colors. The mean $<H-[3.6]>$ color is in the range of predictions from 
dust reddenend \citet[][BC03]{bc03} stellar population synthesis models, but 
the $<[3.6]-[4.5]>$ appears significantly redder by $\sim0.5$ mag.}

No obvious trend of \hone\ color with \hh\ or $[3.6]$ magnitude is seen, but the 
 \hone\ (rest-frame $U_{1700}-B_{4000}$ at $z\sim8$) does appear to correlate
{(at $\sim2\sigma$ confidence)} with \jh\ colors consistent with expectations from dust reddenend 
BC03 stellar population models. {The trend is seen both in IRAC detected galaxies and
in $H_{160}\leq28$ selected samples ($\leq0.3L^*(z=8)$), including IRAC undetected galaxies.}
In contrast, the observed $[3.6]-[4.5]$ colors (rest-frame 
 $B_{4000} - V_{5000}$) are generally redder than the BC03 models, with
 larger offsets at bluer \jh\ colors.

\figcolz

{Figure~3 shows the average spectral energy distributions of IRAC detected galaxies
at $z\sim8$ compared those at $z\sim7$, only 130Myr later.  It is obvious that the observed
$[3.6]-[4.5]$ colors are markedly different. Shifted to the rest-frame, the SEDs viewed together 
suggest a clear flux excess at $\sim5000$\AA, shifting from $[3.6]$ at $z\sim7$ to $[4.5]$ at $z\sim8$.}
The excess is likely caused by [\ion{O}{3}]${4959,5007}$  and $H\beta$ emission lines, 
which are expected to be strong in young star forming galaxies at low metallicities 
(e.g., Schaerer et al. 2010). 

Nevertheless, it is difficult to determine the emission line contribution
using stellar population model fits to the broadband photometry at each redshift separately. 
{Degeneracies  between age and dust in the model colors and systematic uncertainties in implementation of
emission lines hamper such a direct solution. Instead, we will attempt to circumvent this 
limitation by using the joint SED information at the two adjacent redshifts to isolate the 
effective emission line contribution.}

\section{Rest-frame optical emission lines}

Figure~4 shows clearly that the observed 
$[3.6]-[4.5]$ colors at $z\sim8$ are 0.8 mag redder compared to
$z\sim7$, while the $H-[3.6]$ colors are 0.4 mag bluer\footnote{{Note, the samples are 
similar in brightness, $H\sim26.8$ for $z\sim8$ and $H\sim26.5$ for $z\sim7$, hence
both $\sim1.5_{ }L^*(UV)$ in luminosity at $z\sim8$ and $z\sim7$ respectively.}}.
Changes in stellar population  age and/or dust can not produce such differences, 
but strong optical emission lines naturally reproduce the observed change.

With the reasonable assumption that the average rest-frame spectrum is the 
same at both $z\sim8$ and $z\sim7$, we can solve directly for the rest-frame 
equivalent width of these emission lines.

Considering only $[OIII]{\lambda\lambda4959,5007}$
and $H\beta$ and applying the redshift selection functions of \citet{Oesch12} and 
\citet{Bouwens11c}, a linear fit to the colors produces 
W$_{[OIII]{\lambda\lambda4959,5007}+H\beta}=500$\AA. The redshift distribution accounts 
for the fact that emission lines may contribute to both $[3.6]$ and $[4.5]$. Using 
the tabulated emission lines values of Anders \& Fritze-v.Alvensleben (2003) for 
sub-Solar metallicity $0.2Z{_\sun}$ \citep{Erb06,Mai08}, we also correct for the smaller integrated 
contributions of H$\alpha$, H$\beta$, H$\gamma$, [\ion{N}{2}] and [\ion{O}{2}], 
leading to W$_{[OIII]{\lambda\lambda4959,5007}+H\beta}=670$\AA.

Note that this estimate is to first order independent of stellar population models and the fit only 
has one free parameter: the combined rest-frame equivalent width of the strong emission lines. 
The main systematic uncertainty is the redshift selection function for the $z\sim7$ sample,
as the [\ion{O}{3}] line moves from the $[3.6]$ to the $[4.5]$ filter at $z>7.0$
{A secondary uncertainty is the effect of metallicity, which changes the emission line ratios.}
To evaluate the impact of both effects we shift the central redshift of the $z\sim7$ distribution by up to $\Delta z=0.3$ 
in either direction and vary the metallicity from $0.2-1.0_{ }{Z_\sun}$. We add these systematic 
uncertainties in quadrature to the photometric errors arriving at 
W$_{[OIII]\lambda\lambda{4959,5007}+H\beta}=670^{+260}_{-170}$\AA
\footnote{{ Repeating the calculation using a fainter $26.8<H<27.8$ selected subsample, targeting 
galaxies around $L^*(z=8)$ \citep{Oesch12}, yields a similar value
W$_{[OIII]\lambda\lambda{4959,5007}+H\beta}=560$\AA$^{+230}_{-170}$\AA.}}.

The rest-frame equivalent width implies a boost of $0.55^{+0.16}_{-0.11}$ mag to $[4.5]$ 
and smaller $0.19^{+0.10}_{-0.09}$ mag to $[3.6]$ at $z\sim8$, meaning the color of the
stellar continuum is $[3.6]-[4.5]\approx0$. Using the line ratios of Anders \& Fritze-v.Alvensleben 
(2003) for a metallicity of $0.2Z_{\sun}$ and taking into account the slope of the continuum 
($f_\nu\approx$ constant or $f_\lambda\propto\lambda^{-2}$) the corresponding $H\alpha$ equivalent 
is W$_{H\alpha}=430^{+160}_{-110}$. Note that the inferred W$H\alpha$ depends on assuming line
ratios.

\section{specific star formation rates and \\ stellar masses}

{A common approach to estimate the sSFR is derive it from $W_{H\alpha}$, using the 
$H\alpha$ to SFR conversion of \citet{Ke98} and the stellar continuum 
from BC03 stellar populations synthesis models. The model adopts a \citet{Sa55} 
initial mass function (IMF) between $0.1-100$~\msun, a metallicity of 0.2 $Z_\odot$, 
and constant star formation rate. Using the indirect estimate of $H\alpha$ we derive
$sSFR=11^{+11}_{-5}$~Gyr$^{-1}$,
corresponding to an approximate stellar age of $100_{-50}^{+100}$Myr. The estimated sSFR 
is not very sensitive to the assumed star formation history (e.g., declining and rising SFH 
introduce changes $<0.1$ dex), but potentially significant uncertainties
are the intrinsic line ratios,  the possibly larger attenuation of the emission lines 
compared to the continuum and the escape fraction of ionizing photons. Larger attenuation 
and a non-zero escape fraction would increase the derived sSFR.}

Fitting BC03 models with CSF to the broadband SEDs, we inspect the stellar
masses and mass-to-light ratios after correcting the IRAC $[3.6]$ and $[4.5]$ bands 
for emission line contributions. {The derived mass-to-light ratios of IRAC detected
galaxies at $z\sim8$ are $M/L_{1800}\approx0.05$ and $M/L_{V}=0.09$, in good agreement with 
those found by \citet{Ya12} for two bright $z\sim8$ sources.
The $M/L$ would have been overestimated by $\sim3\times$ if emission lines had not been corrected.}
The scatter in $M/L$, based on the intrinsic scatter in $H-[3.6]$ 
and assuming CSF, is $\sigma(M/L_{1800})=0.25$ dex and $\sigma(M/L_V)=0.1$ dex,
where the luminosities refer to the stellar continuum only. This is likely
a lower bound to the true scatter in $M/L$ due to selection effects.
The typical $L*(UV,z=8)$ galaxy (corresponding to $H\sim27.3$) has a median 
rest-frame color $U-V\sim0.2$ (stellar continuum only) and a stellar mass 
$M=0.9_{-0.4}^{+0.6}\times10^{9}~M_\sun$. 

{Model fits to the corrected SEDs can also be used to determine sSFRs, although these 
are rather uncertain due to the degeneracies between age and dust and the weak 
constraints on the UV$-$slope $\beta$ at $z\sim8$ \citep{Bo13}. To mitigate the difficulties somewhat, we can fix 
the reddening to $A_V=0.40\pm0.15$, consistent with the average $\beta$ values
of bright $z\sim7$ galaxies \citep[]{Bo13}. The best-fit sSFR is then $sSFR=7_{-4}^{+13}$~Gyr$^{-1}$,
in good agreement with the sSFR derived from $W_{H\alpha}$.}

We derive integrated stellar mass densities of $H-$band selected galaxies 
to faint UV-limits by converting the stepwise $UV-$luminosity function 
\citep[]{Bouwens11c,Oesch12} to stepwise mass densities, using the best-fit 
$M/L$ to the emission line corrected SED. 

{Note that the emission line correction is derived from 
$\sim 1.5_{ }L^*(UV,z=8)$ galaxies, so we assume that the effect of emission lines 
is similar for $L < L^*$ galaxies. This approximation is unavoidable as sub$-L^*$  galaxies are 
currently beyond the reach of Spitzer, even at our ultradeep limits.} We derive uncertainties by 
randomly drawing $10^4$ LFs, perturbing the stepwise LF and $M/L$ ratios by their 
errors, and integrating the resulting mass densities. Integrating to $M_{UV,AB}=-18$ 
yields $\rho^*(z=8)=0.6^{+0.4}_{-0.3}\times10^6~M_\sun~$Mpc$^{-3}$.

\section{Discussion}
 
This paper presents new ultradeep IRAC data from the IUDF
program  (120 hours in $[3.6]$ and $[4.5]$ over the HUDF)
in combination with a large sample of 63 $Y_{105}-$dropout 
galaxies at  $z\sim8$  to study average HST/WFC3 and Spitzer/IRAC 
SEDs, the effect of emission lines, and their impact
on stellar population model fits.

The most 
{exciting} result is the direct detection of 
{large numbers of} $z\sim8$ galaxies by Spitzer/IRAC {to faint limits}, 
without signs of a hard ``confusion limit''. The ability, for the first time, to determine the
distribution of rest-frame optical colors at $z\sim8$ is a 
{clear} reminder of the enduring capabilities of Spitzer, 
currently in its post cryogenic (``warm'') mission. 

\figewz

Using the joint IRAC constraints on the SED at $z\sim7$ and $z\sim8$, we derive an empirical constraint on 
the emission line contribution of {rest-frame EW W$_{[OIII]4959,5007+H\beta}\sim670$\AA\ 
contributing $\sim0.5$ mag to $[4.5]$ at $z\sim8$. To place this in context,
we compare to recent estimates for star forming galaxies from other surveys (shown in Figure~5). While 
direct comparisons are difficult due to different techniques and 
stellar mass ranges, our results confirm the picture of an increasing 
importance of emission lines towards higher redshift (e.g.,
Shim et al. 2011, Fumagalli et al. 2012, Stark et al. 2013). {The evolution of
the rest-frame equivalent width of $H\alpha$ follows W$_{H\alpha}\propto(1+z)^{1.2\pm0.25}$ 
over $1 < z < 8$. This a factor of $\sim2$ slower than the power law derived 
by Fumagalli et al. (2012) over $0 < z < 2$, but in agreement with 
their model which assumes that the cumulative number density of progenitors at
higher redshift remains constant (e.g., Dokkum et al. 2010, Papovich et al. 2011).}

Clearly, at such high equivalent widths emission lines cannot be ignored when deriving 
stellar masses and ages at high redshift \citep[e.g.,][]{Sch10,deB12}.
After correcting the SEDs for emission lines we find that the 
$L^*(z=8)$ galaxies have stellar masses and $M/L$ that are $\sim3\times$ 
lower. 

\{The $W_{H\alpha}$ based sSFR$=11^{+11}_{-5}$~Gyr$^{-1}$ are $\sim5\times$ higher than
those at $z=2$ (Reddy et al. 2012), consistent with an increase
$sSFR\propto(1+z)^{1.5^{+0.7}_{-0.5}}$ between $2<z<8$. This evolution is 
somewhat higher than derived by Gonzalez et al. (2013) over $2 < z < 6$ 
and somewhat lower than inferred by Stark et al. (2013) at $2 < z < 7$, while in general
agreement with the cumulative number density model of 
Fumagalli et al. 2012.} The latest estimates reduce the tension 
between the observations and simulations, which predicted an evolution 
of the baryonic mass inflow rates  $\propto(1+z)^{2.3}$ 
(e.g., Neistein \& Dekel 2008; Dav\'e et al. 2011, Weinmann et al. 2011).

Obviously, a huge leap in the determination of the 
properties of galaxies at these redshifts will be made possible by
the arrival of JWST. But in the mean time wider and still
deeper future IRAC observations will continue to provide
valuable insight into the properties of the  $z\sim8$ population, 
with emerging detections of $z=9-10$ (e.g., Zheng et al. 2012) 
galaxies probably not far behind.

\acknowledgments

We are grateful to Jarle Brinchmann, Steve Finkelstein, Casey Papovich, Dan Stark, 
Simone Weinmann for helpful discussions. {We thank the anonymous referee for useful
suggestions which improved the paper.} Support for this work was provided by NASA 
through Hubble Fellowship grant HF-51278.01. Support for this work was provided by 
NASA through an award issued by JPL/Caltech. This work has further been supported 
by NASA grant HST-GO-11563.01. We also acknowledge funding from ERC grant 
HIGHZ no. 227749.\\

{\it Facilities:} \facility{Spitzer (IRAC)}, \facility{HST (WFC3/IR)}

\end{document}